\documentclass[conference]{IEEEtran}
\IEEEoverridecommandlockouts
\usepackage{cite}
\usepackage{amsmath,amssymb,amsfonts}
\usepackage{algorithmic}
\usepackage{graphicx}
\usepackage{textcomp}
\usepackage{xcolor}
\def\BibTeX{{\rm B\kern-.05em{\sc i\kern-.025em b}\kern-.08em
    T\kern-.1667em\lower.7ex\hbox{E}\kern-.125emX}}
\begin{document}

\title{A framework for the automation of testing computer vision systems
\thanks{The research described in this paper has been carried out within the FFG Beyond Europe program as part of project 874163 RiSPECT -- Riblet Inspection and Efficiency Assessment Technology funded by the Federal Ministry for Digital and Economic Affairs (BMDW).}
}

\author{\IEEEauthorblockN{Franz Wotawa$^*$\thanks{$^*$Authors are listed in reverse alphabetical order.}}
\IEEEauthorblockA{\textit{Institute for Software Technology} \\
\textit{Graz University of Technology}\\
Graz, Austria \\
wotawa@ist.tugraz.at}
\and
\IEEEauthorblockN{Lorenz Klampfl}
\IEEEauthorblockA{\textit{Christian Doppler Laboratory for}\\
\textit{Quality Assurance Methodologies for}\\
\textit{Autonomous Cyber-Physical Systems}\\
\textit{Institute for Software Technology} \\
\textit{Graz University of Technology}\\
Graz, Austria \\
lklampfl@ist.tugraz.at}
\and
\IEEEauthorblockN{Ledio Jahaj}
\IEEEauthorblockA{\textit{Institute for Software Technology} \\
\textit{Graz University of Technology}\\
Graz, Austria \\
ljahaj@ist.tugraz.at}
}

\maketitle

\begin{abstract}
Vision systems, i.e., systems that allow to detect and track objects in images, have gained substantial importance over the past decades. They are used in quality assurance applications, e.g., for finding surface defects in products during manufacturing, surveillance, but also automated driving, requiring reliable behavior. Interestingly, there is only little work on quality assurance and especially testing of vision systems in general. In this paper, we contribute to the area of testing vision software, and present a framework for the automated generation of tests for systems based on vision and image recognition. The framework makes use of existing libraries allowing to modify original images and to obtain similarities between the original and modified images. We show how such a framework can be used for testing a particular industrial application on identifying defects on riblet surfaces and present preliminary results from the image classification domain.
\end{abstract}


\section{Introduction}

Within the last decades, vision systems and image recognition has gained importance due to its increasing use in practical applications, ranging from automating inspect tasks, surveillance, to autonomous systems including autonomous cars or mobile robots. In all this systems, their behavior heavily rely on the quality of the qualitative or quantitative observations extracted from images. Such vision systems should deliver a correct, i.e., expected output, for the widest range of images. The output should also be as robust as possible. Small changes in the image should not lead to large deviations on side of the output. Especially in case of autonomous driving the latter property has gained a lot of attention. 

\begin{figure}
\begin{center}
\includegraphics[width=0.99\linewidth]{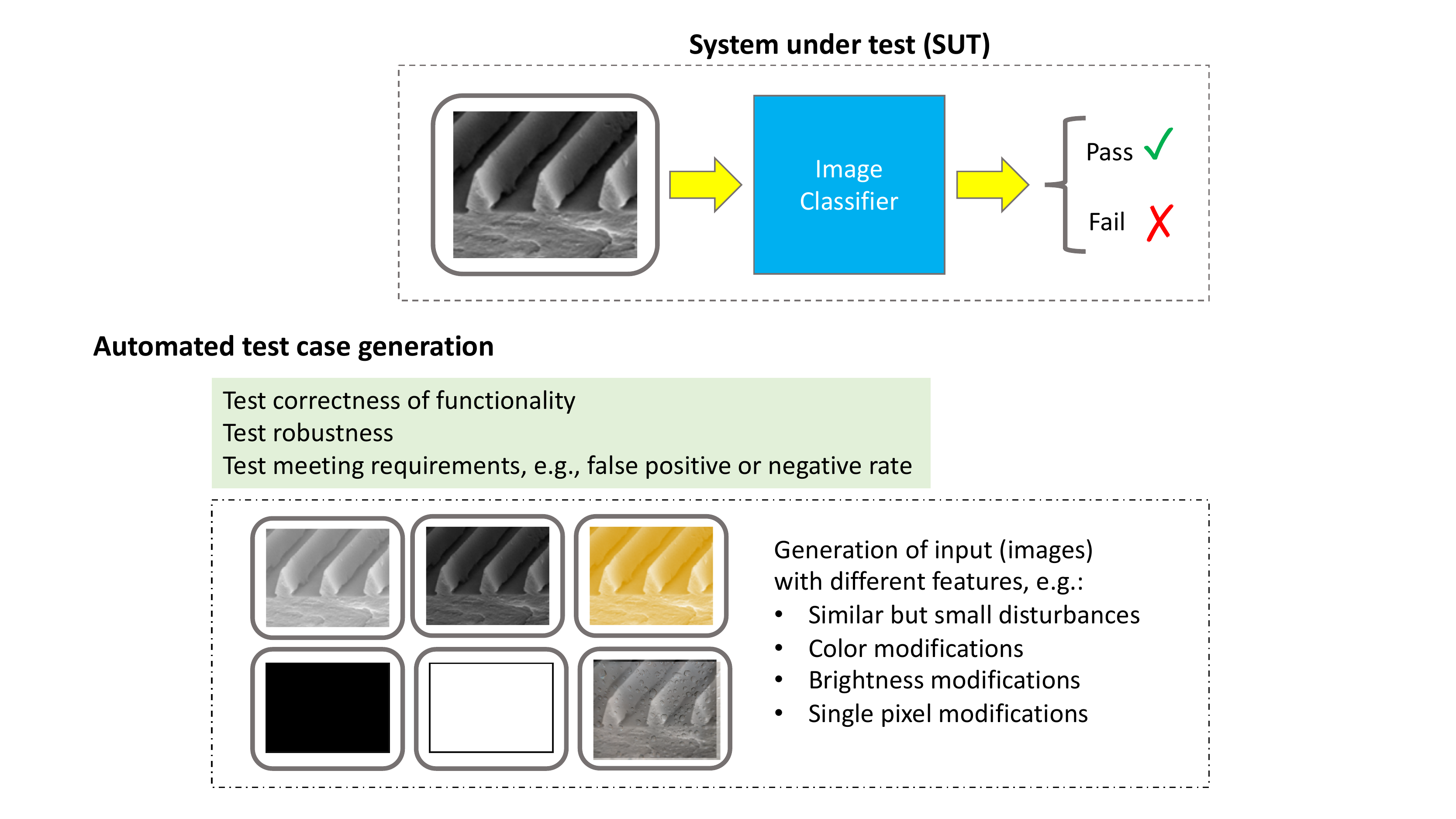}
\end{center}
\caption{The testing approach of the riblet surface quality inspection systems.}
\label{fig:overview}
\end{figure}

Unfortunately, and especially in cases where the vision system is based on the application of machine learning and in particular neural networks, robustness cannot be guaranteed. We may find always an adversarial example, i.e., an image where we change only one or a few pixels, leading to a complete misclassification. There has been a lot of research dealing with testing based on such adversarial examples including~\cite{DBLP:journals/corr/abs-1710-07859} or~\cite{DBLP:conf/cav/HuangKWW17}. In the context of autonomous driving Tian et al.'s work on DeepTest~\cite{DBLP:conf/icse/TianPJR18} has gained great attention. Other similar work include~\cite{DBLP:conf/cvpr/EykholtEF0RXPKS18} and most recently~\cite{DBLP:conf/aitest/ArcainiBBG20}. However, there has also been work on successfully preventing vision systems from being attacked using adversarial examples (see~\cite{DBLP:journals/cacm/GoodfellowMP18}). It is worth noting that testing vision systems does not necessarily only include testing the underlying vision algorithm. Such systems comprise other parts including hardware, which have to work smoothly together for delivering the expected results. Hence, classical testing may be also required (see, e.g.,~\cite{DBLP:conf/icst/NicaJJW17} discussing the use of coverage tools for computer vision applications). 

For our work, we do not rely on a certain underlying method of computer vision used in a particular applications. However, we follow previous research on testing vision software making use of image modifications. The application area of the vision system is the area of surface quality inspection. In particular, we want to test a computer vision application that is currently under development, which returns a quality measure for any riblet surface. Riblets are used for reducing drag of surfaces relying on the shark skin effect, and thus saving energy. Riblets implement tiny structures of a height of about 50$\mu$m. They are used in aviation for reducing fuel consumption. The task of riblet surface quality inspection is to gain information whether the riblet surface is of good or bad quality in order to determine replacement. 

To test the riblet surface quality inspection system, we suggest to generate test cases like depicted in Figure~\ref{fig:overview}. In the approach the system under test will be tested using riblet surface images and also images where we apply modifications. Depending on the modification, the final system should react differently. In case of small deviations, the system should still deliver back the same quality measure or classification. If we change too much, like replacing the original image with a black image, which may happen in practice due to a broken light in the vision system, the system should come up with a corresponding error message indicating the problem. The testing approach we are following does not depend on underlying computer vision methods but is general applicable in this domain. 

In this paper, we present first steps towards implementing the automated test case generation method. This includes a description of how test cases can be generated and a Python framework allowing to make use of different image modifiers, which has been developed to provide a uniform interface for image modifications and which can be easily extended. Moreover, we discuss first preliminary results considering a computer vision based object identifier aiming at showing the practicability of the framework.

\section{Preliminaries}

During development we use a set of images together with their expected output, i.e., a test suite, for coming up with a model used to map images to an expected value. This modeling step can be performed manually or automatically. The latter relies on machine learning approaches like most recently deep neural networks. In any case the initial test suite is also used partially to validate (or evaluate) the model with respect to its capability of performing the correct mapping between images and values.

In testing, however, the aim is not only on validating how good the system under test performs the mapping, i.e., using metrics for quantifying differences between expectations and the real mapping, e.g., the root mean square error, but also to assure that (i) there are no other effects where the system would behave wrong, e.g., finding interactions with the system that cause crashes or exceptions raised during operation, and (ii) the mapping of images to values is robust. The latter is important to guarantee that deviations between images used for obtaining the mapping do not have variations leading to completely wrong values. In addition, we have to test that the system also works well in case of faults in the overall system. For example, due to an image sensors that may break, lenses that may be distorted due to dust and other effects. Such faults should be either detected by the system and lead to warnings or error messages, or being compensated. 

In the following, we formalize the testing problem considering a more general view where we also take care that the system reacts in an expected manner in case of image deviations ranging from smaller deviations to more substantial ones. 

Formally, we state that a system under test $S$ is implementing a function $\theta_S$ mapping an image $\mu_I$ to a value from the domain $D = D_O \cup D_E$, where $D_O$ denotes a set of possible values $S$ returns as a result of image analysis, and $D_E$ a set of other pre-defined behavioral characterizations of $S$ that may be revealed, e.g., a crash or an error message indicating that there is a problem with the image sensor or a lens. Note that $D_O$ may only comprise {\em pass} ($\checkmark$) or {\em fail} ($\times$) in case of an image classifier, but may also comprise elements allowing to state the objects position in the image, as well as an indicator of accuracy of the particular detection. An execution of the system under test always delivers back an element of $D$, i.e., $\theta_S(\mu_I) \in D_O \cup D_E$. 

We define a test case $t$ as a pair $(I,x)$ where $I$ is an image an $x \in D_O$ the expected output. $(I,x)$ is a passing test case if and only if $\theta_S(I) = x$, and a failing test case, otherwise. Furthermore, a test suite $T_S$ for a system under test $S$ is a non-empty set of test cases $t=(I,x)$. As already noted, there is always an initial test suite $T_S^I$ available for a vision system under test $S$ used to come up with a model used for mapping images to a particular domain regardless of the underlying methodology, which may comprise machine learning, and the hardware setup comprising cameras and light sources. 

The testing problem for vision systems $S$, we consider in our work, is to generate test suites from the given system $S$ and the initial test suite $T_S^I$ that assure robustness of the obtained results and the system itself requiring to: (i) come up with slightly modified images from $T_S^I$ where $S$ responses similar, and (ii) other (severe) modifications of images showing that faults in the system like distortions of lenses, or broken sensor or light, are correctly handled by the system. The presented testing framework has been developed considering both types of modifications. 

Formally, the testing framework comprises modification operators $m_1,\ldots,m_k$. The resulting test suite is generated applying all modification operators (and maybe considering all potential parameters) to the initial test suite. Modification operators can be classified as returning similar images or different images where for the latter we assume that the system under test $S$ shall return the error message {\tt err}. We use the predicate {\tt sim} on a modification operator to check for similarity. To solve the testing problem for vision systems, we define the outcome of test suite generation, i.e., a test suite $T_S^\approx$ and $T_S^\epsilon$ for the case of similar images and more sever modifications respectively.
 
$$T_S^\approx = \{(I',x) | (I,x) \in T_S^I \wedge I' = m_i(I) \wedge \mbox{\tt sim($m_i$)} \}$$

$$T_S^\epsilon = \{(I',\mbox{\tt err}) | (I,x) \in T_S^I \wedge I' = m_i(I) \wedge \neg \mbox{\tt sim($m_i$)} \}$$

The testing framework we are going to discuss in more detail in the next section, implements modification operators allowing to come up with both types of test suites.

\section{The testing framework}

We implemented the testing framework for testing computer vision systems in Python 3.8.3. The framework is open, relies on other frameworks that have been integrated, can be easily extended, and provides a simple to use interface to come up with different test suites. Its architecture comprises the following three main parts:

\begin{itemize}
\item {\bf Classifier/Object recognition.} This part allows the integration of a deep neural network to recognize different objects in images. It produces a set of bounding boxes as output and these boxes contain an object each and also their description of what the object is, for example road sign, person, animal, bus, car etc. For our first evaluation we used Mask R-CNN ~\cite{he2018mask} for object detection which was pretrained on the MS COCO dataset ~\cite{lin2015microsoft}.  This part of the framework, takes any image as input and produces the same picture along with the objects recognized inside it. We added this part to the framework for evaluation purposes.

\item {\bf Modifier.} This part implements modification operators allowing to modify any given image. Some of the modification methods we implemented are: inversion, pixel change, affine transformation, blur, add snow/rain/fog/sun, darken, and brighten. All modification operators take an image as input and produce a modification of this image according to the method chosen. After modifying the image, the modifier saves the image into a directory/folder specified by the user.

\item {\bf Image diff.}  We implemented a method for finding the difference between two given images. The diff tool uses an algorithm called Structural Similarity Index Measure (SSIM)~\cite{1284395}, which is used to compute the similarity between two images. It takes into account the structural information of the image and compares it accordingly. In contrast to other techniques like mean squared error (MSE), SSIM finds a pattern and relation of pixels of an image and defines a structure. Afterwards, it compares the structures of two images. SSIM returns a value between 0 and 1 where 0 means that two images are completely different, and 1 means the two images are identical. 
\end{itemize}

The framework provides a command line interface and is very simple to use. The user can classify any image that he or she prefers, can modify images, and also compare two images. The framework also allows to implement a test case generator as described in the previous section.

In the following, we provide an initial experimental study making use of the implemented R-CNN. 

\section{Preliminary experimental study}

\begin{figure*}[t]
\begin{center}
\begin{tabular}{cccc}
\hline
Original {\bf Image 1} & colors inverted & black pixels added & \\ 
\includegraphics[width=3.4cm]{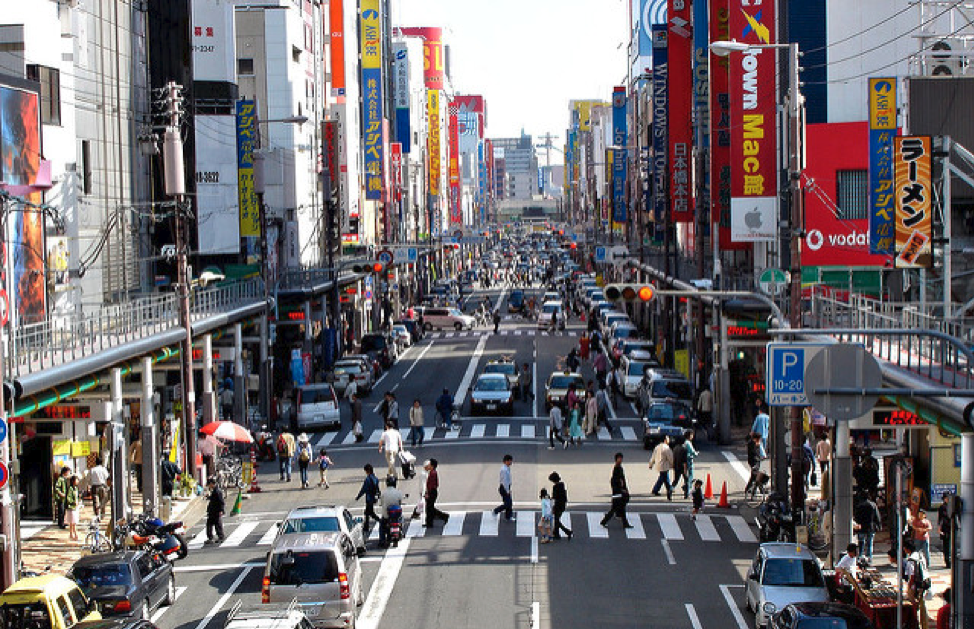}  & \includegraphics[width=3.4cm]{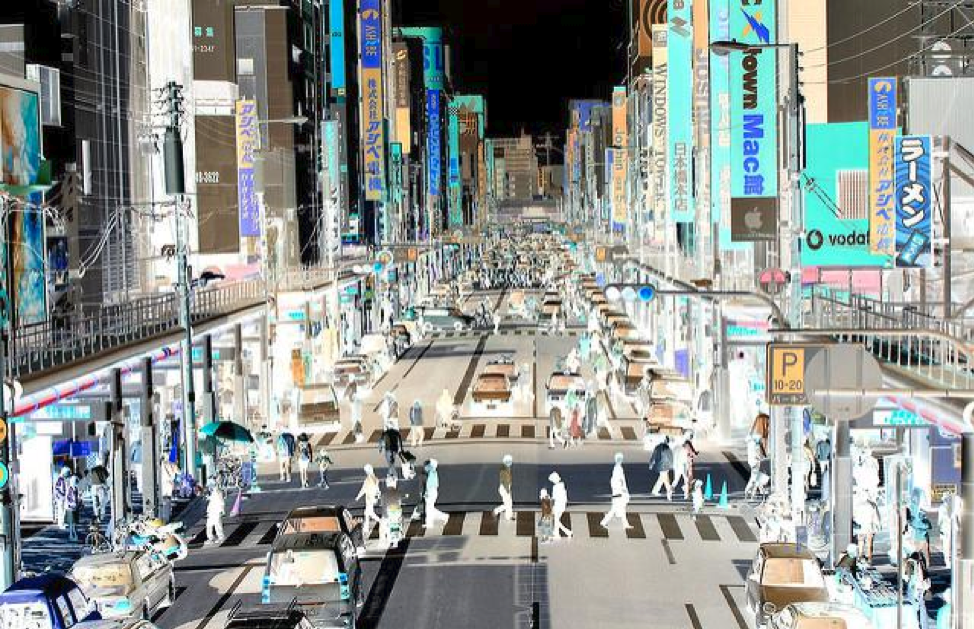}  & \includegraphics[width=3.4cm]{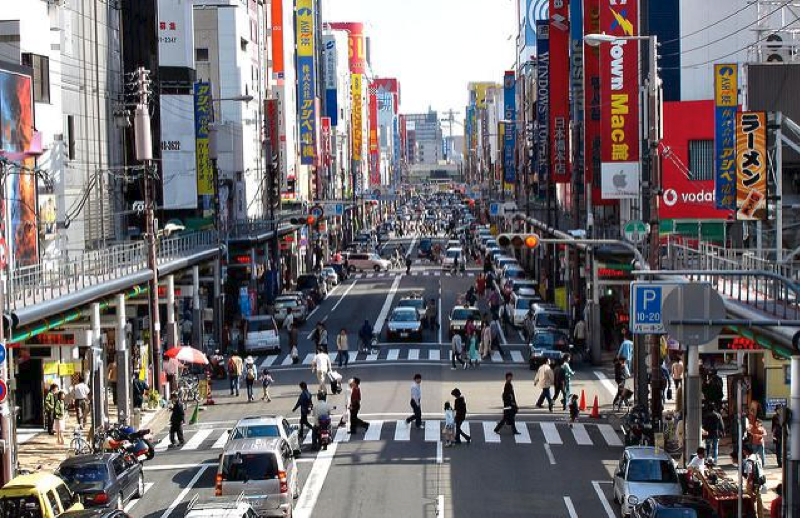} & \\
\hline
Original {\bf Image 2} & 80\% blurring & affine transformation & \\ 
\includegraphics[width=3.4cm]{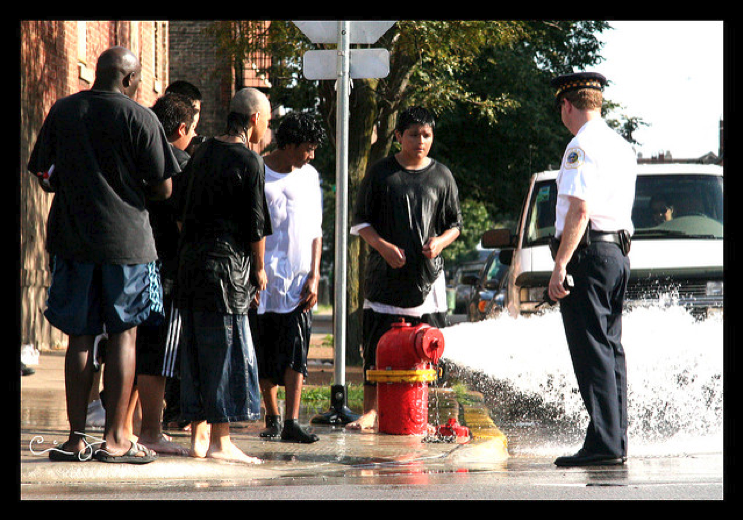}  & \includegraphics[width=3.4cm]{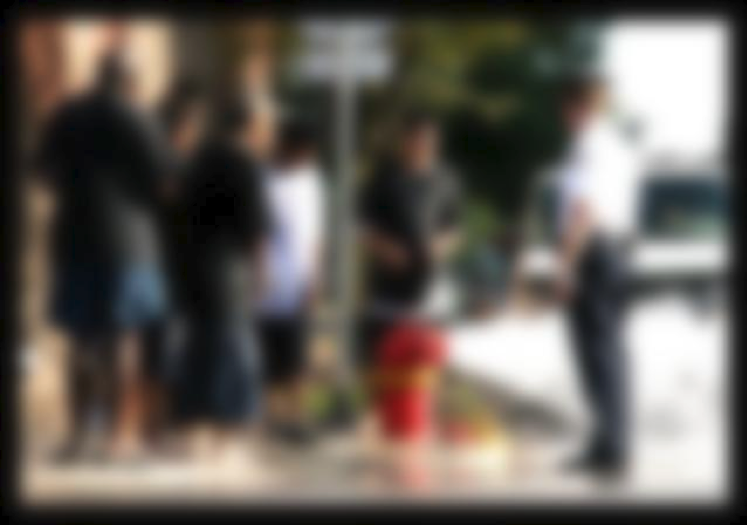}  & \includegraphics[width=3.4cm]{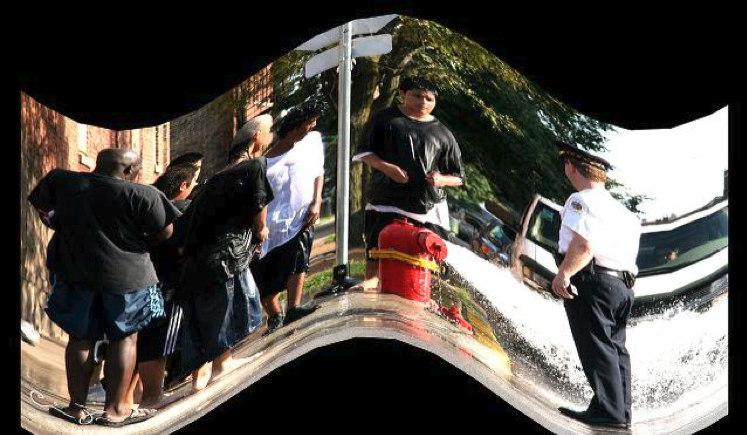} & \\
\hline
Original {\bf Image 3} & darkened & flipped & \\ 
\includegraphics[width=3.4cm]{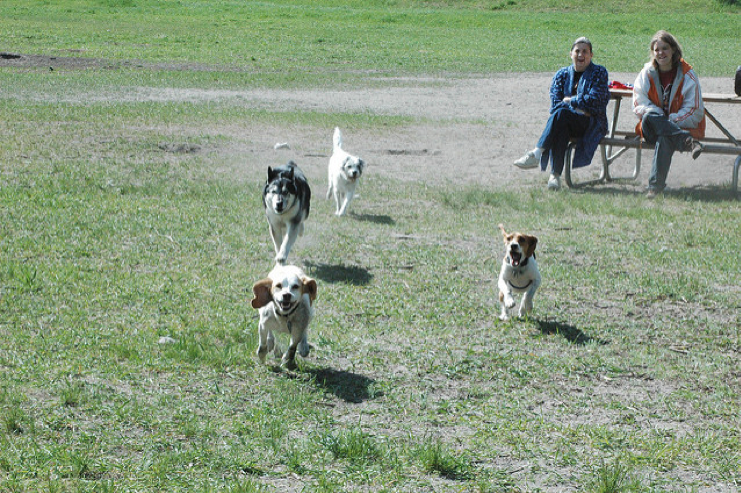}  & \includegraphics[width=3.4cm]{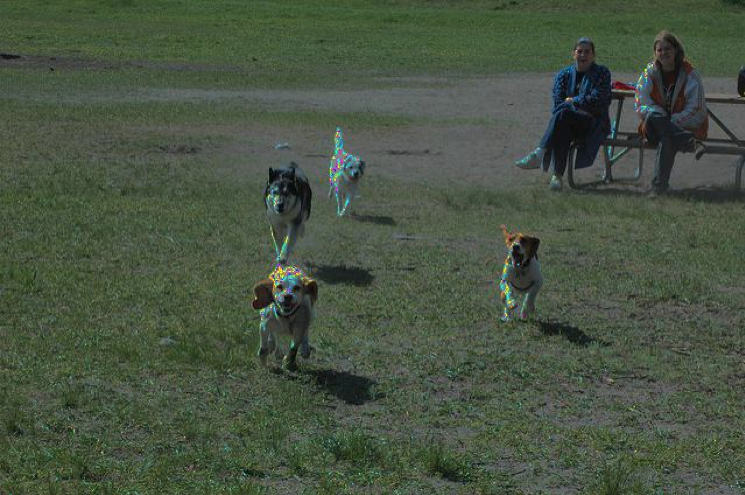}  & \includegraphics[width=3.4cm]{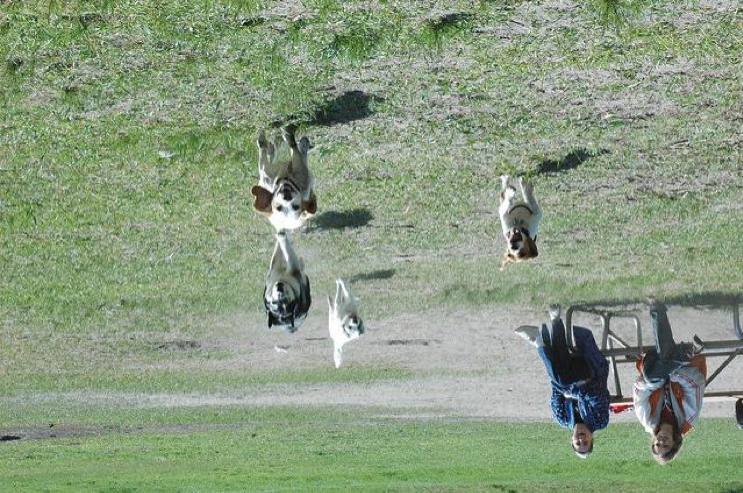} & \\
\hline
Original {\bf Image 4} & brightened & fog added & shadows added \\ 
\includegraphics[width=3.4cm]{images/image4_orig.png}  & \includegraphics[width=3.4cm]{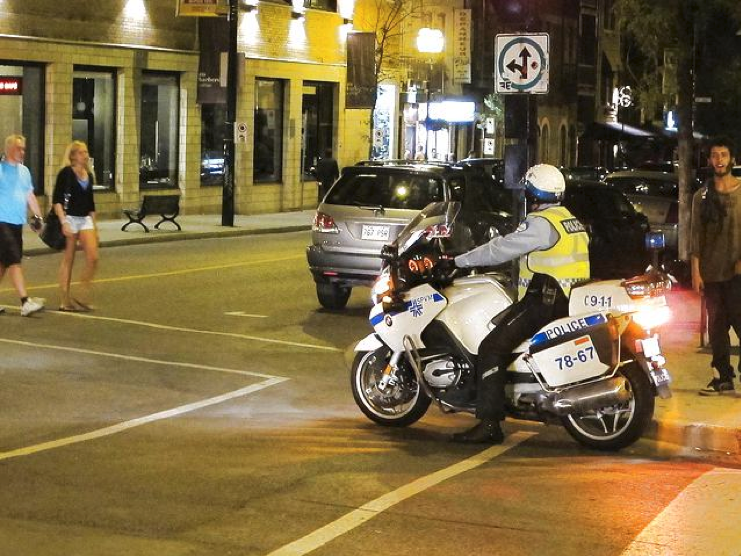}  & \includegraphics[width=3.4cm]{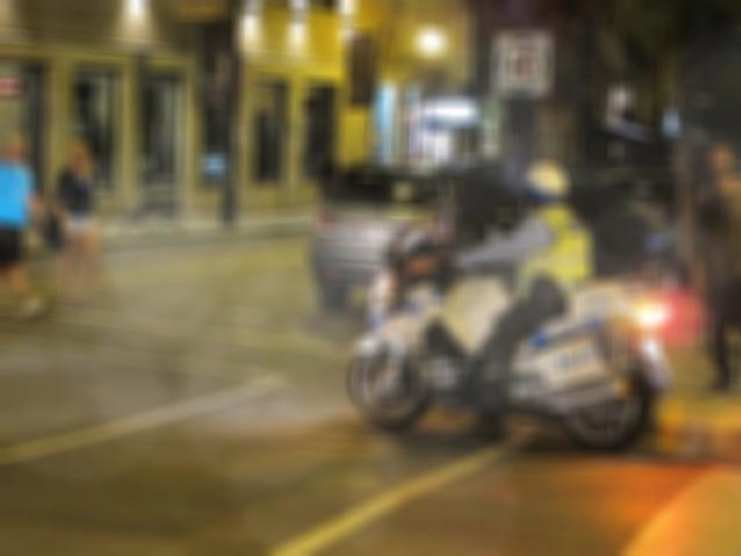} &  \includegraphics[width=3.4cm]{images/image4_shadows.png} \\
\hline
Original {\bf Image 5} & snow added & rain added & sun added \\ 
\includegraphics[width=3.4cm]{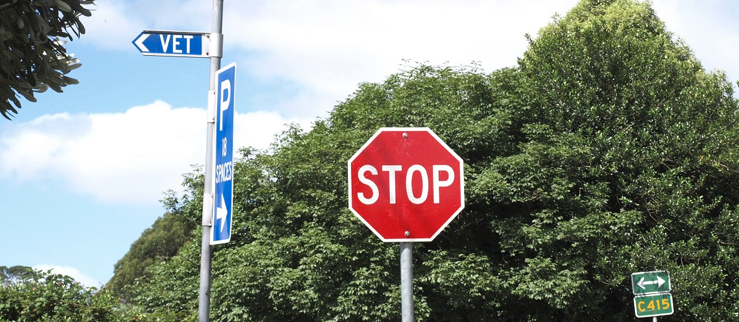}  & \includegraphics[width=3.4cm]{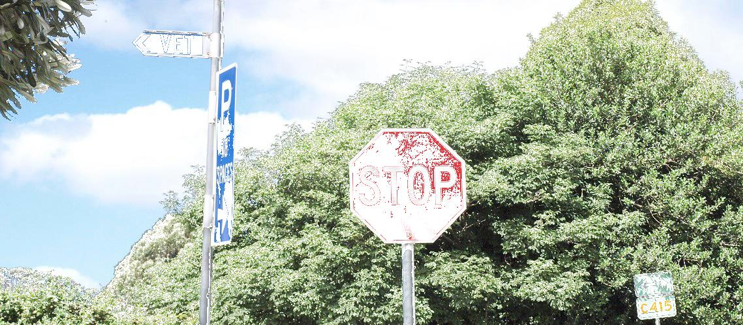}  & \includegraphics[width=3.4cm]{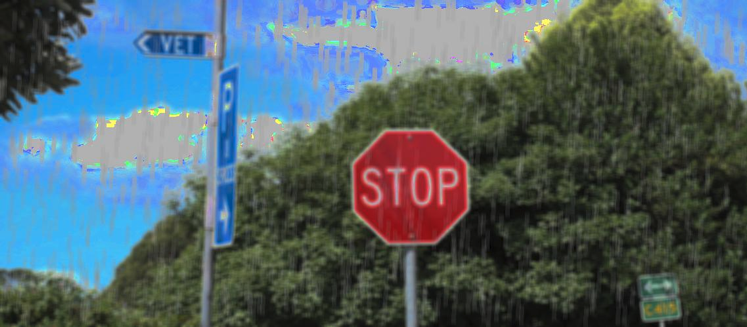} &  \includegraphics[width=3.4cm]{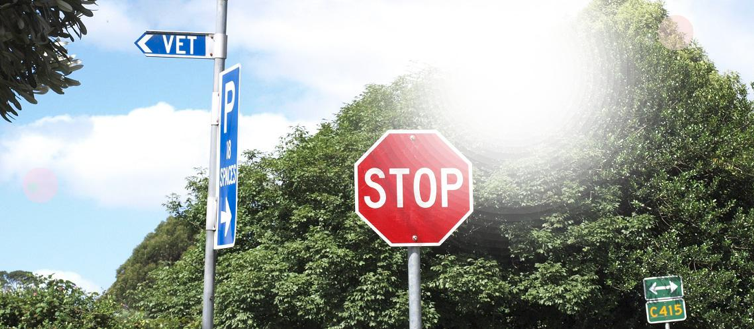} \\
\hline
\end{tabular}
\end{center}
\caption{The images used for our experiments.}
\label{fig:images}
\end{figure*}

The objective behind the experimental study was to indicate the usefulness of the proposed testing framework for computer vision applications. Due to the fact that the riblet surface inspection system is currently under development and not ready for testing, we made use of the R-CNN, which is part of our framework. In our experiments, we considered five images and applied several different modification methods. We depict the original images and the considered modifications in Figure~\ref{fig:images}.

After applying the modification operators, we computed the SSIM difference between the original image and the modified one, and used the R-CNN on the modified images. The obtained outcome we summarize in Table \ref{tab:ssim}.

\begin{center}
\begin{table}
\caption{Computed SSIM values after the applied image modifications and object recognition results.}
\begin{tabular}{llll}
\hline
 & \bf modification & \bf SSIM & \bf object recognition results  \\
\hline
1 & inverted & 0.50 & \parbox{4.5cm}{Some objects like cars are misclassified as boats.} \\[1em]
1 & pixels & 0.93 & \parbox{4.5cm}{Most of the classifications are correct but objects far away are not recognized anymore.} \\[1em]
\hline 
2 & blur & 0.27 & \parbox{4.5cm}{Classification correct except for cars and the policeman.} \\[1em]
2 & affine & 0.00 & \parbox{4.5cm}{Classification correct except for one car recognized as boat. } \\[1em]
\hline
3 & darkened & 0.55 & \parbox{4.5cm}{One of the dogs recognized as a kite.} \\[1em]
3 & flipped & 0.05 & \parbox{4.5cm}{Only one object recognized correctly.} \\[1em]
\hline
4 & brightened & 0.84 &Classification correct. \\ 
4 & fog & 0.41  & Classification correct. \\
4 & shadow & 0.87 & Classification correct.\\
\hline
5 & snow & 0.60 & Classification correct. \\
5 & rain & 0.34 & Classification correct. \\
5 & sun & 0.82 & Classification correct. \\
\hline
\end{tabular}
\label{tab:ssim}
\end{table}
\end{center}

From the results, we see that in many cases the R-CNN works fine only failing for some objects. We also see that there seems to be little correlation between the SSIM and the degradation of object recognition. For example, when applying fog to image 4 causing the SSIM to be 0.41 the recognition of objects is not influenced. Almost the same happens for image 2 with affine transformation, resulting in only one wrongly recognized object but having a SSIM of 0.00. Therefore, further investigations are required to clarify the relationship between SSIM, the modification operator used, and the classification results. 

\section{Conclusions}

In this paper, we introduced a framework for the automated generation of tests for systems utilizing vision and image recognition since often testing is only performed by evaluating the system on a image test set without specific modifications. In particular, the framework architecture comprises three parts, i.e., the classifier or object recognition part which allows the integration of different deep neural networks used for image recognition which should be validated, the modifier part which applies various image modification and that can easily be extended by additional modifications and a part for calculating the differences between the original image and the modified image based on a structural similarity index measure metric which can also be extended by additional metrics of interest.

From the first results it can be seen that there seems to be little correlation between the applied metric and the recognition of objects. Therefore, we intend to further work on the integration of additional metrics as well as the integration of different object detectors to investigate if this is the case in general. Moreover, we may consider also the applied modification operator among other parameters when investigating on the relationship between metrics and object recognition results.
In future work, we will make use of the framework for testing the mentioned riblet surface inspection system, which is currently under development.

%

\end{document}